\documentclass[aps,prev,twocolumn,preprintnumbers,floatfix,nofootinbib]{revtex4-1}
\pdfoutput=1
\usepackage[english]{babel}
\usepackage{graphicx,color}
\usepackage{bm}
\usepackage{times}
\usepackage{slashed}
\usepackage{multirow}
\usepackage[usenames,dvipsnames,svgnames,table]{xcolor}
\usepackage{adjustbox}
\usepackage{slashed}
\usepackage{booktabs}
\usepackage{makecell}

\usepackage{amsthm}
\usepackage{amssymb,amsmath}
\usepackage{subfigure}
\usepackage{hyperref}
\usepackage[dvipsnames]{xcolor}
\definecolor{bluencs}{rgb}{0.0, 0.53, 0.74}
\definecolor{darkcyan}{rgb}{0.0, 0.55, 0.55}
\definecolor{hanblue}{rgb}{0.27, 0.42, 0.81}
\hypersetup{
	linktocpage=true,
	setpagesize=true,
	urlcolor=blue,
	citecolor=blue,
	linkcolor=blue, 
	menucolor=cyan,
	colorlinks=true, 
	filecolor=magenta,
	citebordercolor=blue,
	pdftitle={Search for Charged Higgs Bosons through Vector-Like Top Quark Pair Production at the LHC},
	pdfsubject={Latex},
	pdfauthor={Bk},
	pdfkeywords={BSM},
	unicode=true
}
\usepackage[capitalize]{cleveref}

\newcommand{\be}{\begin{equation}}
\newcommand{\ee}{\end{equation}}
\newcommand{\bea}{\begin{eqnarray}}
\newcommand{\eea}{\end{eqnarray}}

\usepackage{xfrac}
 
\usepackage{float}
\usepackage{bbding,pifont}
\definecolor{brilliantrose}{rgb}{1.0, 0.33, 0.64}
\definecolor{lawngreen}{rgb}{0.49, 0.99, 0.0}
\definecolor{magenta}{rgb}{1.0, 0.0, 1.0}
\makeatletter
\let\old@float\@float
\def\@float{\let\centering\relax\old@float}
\makeatother
\begin{document}

\title{Search for charged Higgs bosons through vector-like top quark pair production at the LHC}

\author{A. Arhrib$^{1,2}$}
\email{aarhrib@gmail.com}
\author{R. Benbrik$^{3}$}
\email{r.benbrik@uca.ac.ma}
\author{M. Berrouj$^{3}$}
\email{mbark.berrouj@ced.uca.ma}
\author{M. Boukidi$^{3,4}$}
\email{mohammed.boukidi@ced.uca.ma}
\author{B. Manaut$^{5}$}
\email{b.manaut@usms.ma}

\affiliation{$^1$ Universit´e Abdelmalek Essaadi, Faculty of Sciences and Techniques, B. 416, Tangier, Morocco.}
\affiliation{$^2$  Department of Physics and Center for Theory and Computation, National Tsing Hua University,
	Hsinchu, Taiwan 300.}

\affiliation{$^3$ Laboratory of Physics, Energy, Environment, and Applications, Cadi Ayyad University,
	Sidi Bouzid, P.O. Box 4162, Safi, Morocco.}
\affiliation{$^4$ Institute of Nuclear Physics, Polish Academy of Sciences, ul. Radzikowskiego 152, Cracow, 31-342, Poland.}
\affiliation{$^5$ Polydisciplinary Faculty, Laboratory of Research in Physics and Engineering Sciences, Team of Modern and Applied Physics, Sultan Moulay Slimane University, Beni Mellal, Morocco.}

\begin{abstract}
	We investigate the discovery prospects for a vector-like top partner (VLT) in the Type-II Two-Higgs-Doublet Model (2HDM-II) extended by a vector-like quark doublet ($TB$) at the 14~TeV LHC. The study focuses on the pair production process $pp \to T\bar{T} \to bH^+\, \bar{b}H^- \to b(tb)\, \bar{b}(\bar{t}b)$, yielding fully hadronic final states characterized by high $b$-jet multiplicity. Two analysis strategies are employed, requiring at least four or five $b$-tagged jets (4$b$ and 5$b$), to exploit the signal topology. Assuming a charged Higgs mass of $m_{H^\pm} \sim 600$~GeV and a systematic uncertainty of $\delta = 5\%$, the 4$b$ channel enables discovery up to $m_T \sim 1200$~GeV at $\mathcal{L} = 300~\text{fb}^{-1}$, while the 5$b$ analysis extends the reach to $m_T \sim 1300$~GeV. At higher luminosities of 1000--3000~fb$^{-1}$, the 5$b$ strategy achieves discovery sensitivity up to $m_T \sim 1400$~GeV. The reach is significantly reduced as $m_{H^\pm}$ increases, due to the suppression of the $\text{BR}(T \to H^+ b)$: for $m_{H^\pm} \sim 1000$~GeV, discovery becomes unattainable across all luminosity and systematic uncertainty configurations. Sensitivity is also strongly impacted by systematic uncertainties: for $\delta = 10\%$ and $m_{H^\pm} < 1000$~GeV, discovery remains viable up to $m_T \sim 1200$~GeV in the 5$b$ analysis, while for $\delta = 20\%$, no discovery is achievable for $m_T \geq 1000$~GeV. 
\end{abstract}

\maketitle

\section{Introduction}

The discovery of the Higgs boson in 2012 by the ATLAS and CMS collaborations at the Large Hadron Collider (LHC) confirmed the Higgs mechanism as the cornerstone of mass generation in the Standard Model (SM)~\cite{ATLAS:2012yve, CMS:2012qbp}. While this milestone solidified the SM framework, it also left open questions about the scalar sector, sparking interest in extensions of the SM. Among the most studied extensions, the Two-Higgs-Doublet Model (2HDM)~\cite{Branco:2011iw, Draper:2020tyq} introduces an additional scalar doublet, expanding the Higgs sector with neutral ($H$, $A$) and charged ($H^\pm$) bosons. These new particles offer a rich phenomenology and provide pathways for exploring signatures of new physics at current and future colliders.

One compelling extension within the 2HDM framework is the incorporation of vector-like quarks (VLQs) \cite{Aguilar-Saavedra:2013qpa,Buchkremer:2013bha,Fuks:2016ftf,Alves:2023ufm,Gopalakrishna:2013hua, Han:2023ied,Han:2025itd,Han:2022jcp,Yang:2022wfa,Tian:2021oey,Yang:2021dtc,Cao:2022mif,Yang:2024aav,Banerjee:2024zvg,Banerjee:2023upj,Vignaroli:2012si,Vignaroli:2015ama}, heavy fermions with identical quantum numbers for their left- and right-handed components. Unlike their SM counterparts, VLQs couple to the Higgs sector through Yukawa interactions that are independent of electroweak symmetry breaking. These particles, classified as singlets, doublets, or triplets under $SU(2)_L$, often arise in theoretical frameworks such as Randall-Sundrum models~\cite{Randall:1999ee, Carena:2007tn, Gopalakrishna:2011ef}, $E_6$-based Grand Unified Theories (GUTs)~\cite{Hewett:1988xc}, Little Higgs models~\cite{Arkani-Hamed:2002ikv, Schmaltz:2005ky}, and composite Higgs scenarios~\cite{Dobrescu:1997nm, Hill:2002ap, Agashe:2004rs, Barbieri:2007bh}. They also feature prominently in extensions of the 2HDM, offering new decay channels and collider signatures~\cite{Arhrib:2024tzm,Arhrib:2024dou, Arhrib:2024mbq, Benbrik:2022kpo,Abouabid:2023mbu,Benbrik:2023xlo,  Benbrik:2022kpo,Benbrik:2024bxt, Aguilar-Saavedra:2017giu, Dermisek:2019vkc,Dermisek:2021zjd,Dermisek:2020gbr,Ghosh:2023xhs}.

In the 2HDM framework with VLQs, novel decay channels such as $T/B \to H^\pm b/t$, $T/B \to H t/b$, and $T/B \to A t/b$ emerge, substantially altering the branching ratios of VLQs. This modifies the constraints derived from searches traditionally focused on SM-like decay modes ($Wq$, $Zq$, $hq$). As a result, scenarios involving VLQs coupled to an extended Higgs sector, such as the 2HDM + VLQs, provide distinctive collider signatures that merit dedicated investigation.

Experimental efforts by ATLAS and CMS have extensively probed VLQ production through single and pair production mechanisms~\cite{ATLAS:2024fdw, CMS:2024bni}. While no experimental evidence has been found so far, this may suggest that VLQs preferentially decay into non-SM particles, which have not been the focus of current searches. This observation highlights the importance of exploring scenarios like 2HDM + VLQ, where VLQs decay into additional Higgs bosons, potentially leading to unexplored and distinctive experimental signatures.

A previous study~\cite{Dermisek:2020gbr} investigated similar VLQ decay modes in the 2HDM+VLQ framework, focusing on 95\% CL exclusion limits. In contrast, this work aims to assess the discovery potential of VLQs decaying via charged Higgs bosons, extending beyond exclusion-based analyses. Specifically, we examine the pair production of  a vector-like top partner (VLT) in the $TB$ doublet scenario, where the dominant decay mode is $T \to H^+ b$, followed by $H^+ \to tb$. This process leads to a distinctive multi-$b$-jet signature, offering a viable search channel at the LHC. Our study  explores different charged Higgs mass scenarios (600, 800, and 1000 GeV) and accounts for background uncertainties to ensure a realistic assessment of the discovery prospects. The results highlight the potential sensitivity of this channel across a broad range of model parameters, motivating further exploration at the LHC and future colliders.

The paper is organized as follows: In Section~\ref{sec-F}, we introduce the 2HDM-II+VLQs framework and outline the simulation setup for production and decay processes. Section~\ref{sec-A} presents the numerical results. Finally, Section~\ref{sec:conclusion} summarizes our findings and discusses their implications for future collider experiments.

\section{Framework}
\label{sec-F}
This section provides a concise overview of the 2HDM-II+VLQ framework. We begin by revisiting the well-known $\mathcal{CP}$-conserving scalar potential of the 2HDM, involving two Higgs doublets $(\Phi_1, \Phi_2)$ subject to a softly broken discrete $\mathbb{Z}_2$ symmetry, $\Phi_1 \to -\Phi_1$, which is softly violated by dimension-2 terms \cite{Branco:2011iw, Gunion:1989we}:

\begin{eqnarray} \label{pot}
\mathcal{V} &=& m^2_{11}\Phi_1^\dagger\Phi_1+m^2_{22}\Phi_2^\dagger\Phi_2
-\left(m^2_{12}\Phi_1^\dagger\Phi_2+{\rm h.c.}\right)
\nonumber \\
&&+\frac{1}{2}\lambda_1\left(\Phi_1^\dagger\Phi_1\right)^2
+\frac{1}{2}\lambda_2\left(\Phi_2^\dagger\Phi_2\right)^2 \nonumber \\
&& \qquad +\lambda_3\Phi_1^\dagger\Phi_1\Phi_2^\dagger\Phi_2
+\lambda_4\Phi_1^\dagger\Phi_2\Phi_2^\dagger\Phi_1
\nonumber \\
&&+\left[\frac{1}{2}\lambda_5\left(\Phi_1^\dagger\Phi_2\right)^2+{\rm h.c.}\right].
\end{eqnarray}
All parameters in this potential are real. The two complex scalar doublets $\Phi_{1,2}$ can be rotated into a basis, $H_{1,2}$, where only one of them acquires a Vacuum Expectation Value (VEV). By employing the minimization conditions of the potential for EWSB, the 2HDM can be parametrized by seven independent quantities: $m_h$, $m_H$, $m_A$, $m_{H^\pm}$, $\tan\beta = v_2/v_1$, $\sin(\beta - \alpha)$, and the soft-breaking parameter $m^2_{12}$

To suppress tree-level Flavor Changing Neutral Currents (FCNCs), the 2HDM admits four distinct Yukawa configurations, depending on how the $\mathbb{Z}_2$ symmetry is extended to the fermion sector\footnote{This work specifically focuses on the Type-II 2HDM within the alignment limit \cite{Draper:2020tyq}, where the lightest neutral Higgs boson is identified with the discovered 125 GeV state.}. These configurations are: Type-I, where $\Phi_2$ couples to all fermions, Type-II, where $\Phi_2$ couples to up-type quarks and $\Phi_1$ to down-type quarks and charged leptons, Type-Y (Flipped), where $\Phi_2$ couples to up-type quarks and charged leptons, and $\Phi_1$ to down-type quarks, and Type-X (Lepton Specific), where $\Phi_2$ couples to quarks, and $\Phi_1$ to charged leptons.

Next, we introduce the VLQ component of the model. The gauge-invariant interactions between the new VLQs and SM particles arise from renormalizable couplings, with the VLQ representations specified as follows:
\begin{align}
& T_{L,R}^0 \, && \text{(singlets)} \,, \notag \\
& (X\,T^0)_{L,R} \,, \quad (T^0\,B^0)_{L,R} \, && \text{(doublets)} \,, \notag \\
& (X\,T^0\,B^0)_{L,R} \,, \quad (T^0\,B^0\,Y)_{L,R}  && \text{(triplets)} \,.
\end{align}

In this context, the superscript zero distinguishes weak eigenstates from mass eigenstates. The electric charges of the VLQs are $Q_T = 2/3$, $Q_B = -1/3$, $Q_X = 5/3$, and $Q_Y = -4/3$, with $T$ and $B$ sharing the same electric charges as the SM top and bottom quarks, respectively.

The physical up-type quark mass eigenstates may, in general, contain non-zero $Q_{L,R}^0$ components (where $Q$ represents the VLQ field) when new fields $T_{L,R}^0$ and $B_{L,R}^0$ with non-standard isospin assignments are added to the SM. This scenario leads to deviations in their couplings to the $Z$ boson. These deviations are constrained by atomic parity violation experiments and measurements of $R_c$ at LEP \cite{ParticleDataGroup:2012pjm}, which impose stringent limits for the up and charm quarks but are comparatively less restrictive for the top quark.

In the $\mathbb{Z}_2$ basis, the Yukawa Lagrangian for the three SM quark generations $q_{Li} = (u^0_{Li} \; d^0_{Li})^T$ ($i = 1, 2, 3$) and a VLQ doublet $Q_{L,R} = (T^0 \; B^0)^T_{L,R}$ is given by:
\begin{eqnarray}
\mathcal{L}_Y &=& - y_{ij}^u \bar{q}_{Li} \tilde{\Phi}_2 u_{Rj} - y_{ij}^d \bar{q}_{Li} \Phi_1 d_{Rj} \nonumber\\&&- y_{4j}^u \bar{Q}_L \tilde{\Phi}_2 u_{Rj} - y_{4j}^d \bar{Q}_L \Phi_1 d_{Rj} + \text{h.c.},
\end{eqnarray}

where $\Phi_1 = (H_d^+ \; H_d^0)^T$ and $\Phi_2 = (H_u^+ \; H_u^0)^T$ are the Higgs doublets, and $\tilde{\Phi}_2 = i \sigma_2 \Phi_2^*$ is the conjugate Higgs doublet. The Yukawa couplings $y_{ij}^{u,d}$ describe the interactions between the SM quarks and the Higgs doublets, where $i, j = 1, 2, 3$. The couplings $y_{4j}^{u,d}$ represent the interactions between the VLQs and the SM quarks.

We assume that the VLQ doublet eigenstates dominantly mix with the third generation, as motivated by the natural mass hierarchy~\cite{Aguilar-Saavedra:2013wba} and consistent with stringent experimental constraints from low-energy flavor-changing processes~\cite{Barger:1995dd, Frampton:1999xi, Barenboim:2001fd, Aguilar-Saavedra:2002phh}. Therefore, the mixing with the first two generations can be neglected, and we set $y_{41}^{u,d} = y_{42}^{u,d} = 0$, retaining only $y_{43}^{u,d} \neq 0$.

The mixing between the third-generation SM quarks and the VLQs is governed by $2 \times 2$ unitary matrices $U_{L,R}^u$ and $U_{L,R}^d$, which relate the weak eigenstates to the mass eigenstates. For the up-type quarks $t$ and $T$:

\begin{equation}
\begin{pmatrix} t_{L,R} \\ T_{L,R} \end{pmatrix} = \begin{pmatrix} \cos \theta_{L,R}^u & -\sin \theta_{L,R}^u e^{i \phi_u} \\ \sin \theta_{L,R}^u e^{-i \phi_u} & \cos \theta_{L,R}^u \end{pmatrix} \begin{pmatrix} t_{L,R}^0 \\ T_{L,R}^0 \end{pmatrix},
\end{equation}

and for the down-type quarks\footnote{Measurements of $R_b$ at LEP set constraints on the $b$ mixing with the new fields that are stronger than for mixing with the lighter quarks $d,s$ \cite{Aguilar-Saavedra:2002phh}. } $b$ and $B$:

\begin{equation}
\begin{pmatrix} b_{L,R} \\ B_{L,R} \end{pmatrix} = \begin{pmatrix} \cos \theta_{L,R}^d & -\sin \theta_{L,R}^d e^{i \phi_d} \\ \sin \theta_{L,R}^d e^{-i \phi_d} & \cos \theta_{L,R}^d \end{pmatrix} \begin{pmatrix} b_{L,R}^0 \\ B_{L,R}^0 \end{pmatrix}.
\end{equation}

The mass terms for these quarks in the weak eigenstate basis are:

\begin{eqnarray}
\mathcal{L}_\text{mass} &=& - \begin{pmatrix} \bar{t}_L^0 & \bar{T}_L^0 \end{pmatrix} \begin{pmatrix} y_{33}^u \frac{v}{\sqrt{2}} & 0 \\ y_{43}^u \frac{v}{\sqrt{2}} & M^0 \end{pmatrix} \begin{pmatrix} t_R^0 \\ T_R^0 \end{pmatrix} \nonumber\\&&- \begin{pmatrix} \bar{b}_L^0 & \bar{B}_L^0 \end{pmatrix} \begin{pmatrix} y_{33}^d \frac{v}{\sqrt{2}} & 0 \\ y_{43}^d \frac{v}{\sqrt{2}} & M^0 \end{pmatrix} \begin{pmatrix} b_R^0 \\ B_R^0 \end{pmatrix} + \text{h.c.}.\label{eq:L_mass}
\end{eqnarray}
where $v = 246$ GeV is the Higgs vacuum expectation value (VEV), and $M^0$ is the bare VLQ mass term\footnote{This gauge-invariant mass term is independent of the Higgs mechanism. It can either appear directly in the Lagrangian or arise from a Yukawa coupling to a scalar singlet that acquires a large VEV $v' \gg v$.}.

In the $TB$ doublet scenario, gauge invariance ensures that the off-diagonal elements $y_{34}^{u,d}$ vanish, leaving $y_{33}^{u,d}$ and $y_{43}^{u,d}$ as the only non-zero Yukawa couplings. The mass matrix is diagonalized by the mixing matrices $U_L^q$ and $U_R^q$, which satisfy:
\begin{equation}
U_L^q \, \mathcal{M}^q \, (U_R^q)^\dagger = \mathcal{M}^q_\text{diag} \,, 
\label{ec:diag}
\end{equation}
where $\mathcal{M}^q$ represents the mass matrices in Eq.~(\ref{eq:L_mass}), and $\mathcal{M}^q_\text{diag}$ are their diagonalized forms.

The left- and right-handed mixing angles for the quarks are related by:
\begin{equation}
\tan \theta_L^u \approx \frac{m_t}{m_T} \tan \theta_R^u, \quad \tan \theta_L^d \approx \frac{m_b}{m_B} \tan \theta_R^d,
\end{equation}

where $m_t$ and $m_b$ are the masses of the SM top and bottom quarks, and $m_T$ and $m_B$ are the VLQ masses.

The $TB$ doublet scenario within the 2HDM-II framework provides a compelling avenue for exploring charged Higgs bosons through VLQ decays. In this setup, the VLT decays into a charged Higgs boson ($H^\pm$) and a bottom quark ($b$) with a substantial branching ratio~\cite{Benbrik:2022kpo}. The decay width\footnote{Detailed analytical expressions for the couplings are provided in the Appendix of Ref.~\cite{Benbrik:2022kpo}.} for the process $T \to H^+ b$ is given by:

\begin{align}
\Gamma(T \to H^+ b) & = \frac{g^2 }{64 \pi} 
\frac{m_T}{M_W^2} \lambda(m_T,m_b,M_{H^\pm})^{1/2}
\nonumber \\
&\times\left\{ (|Z_{Tb}^L|^2 \cot^2 \beta + |Z_{Tb}^R|^2 \tan^2 \beta )\right. \notag \\
& \left. \times  \left[1+r_b^2 - r_{H^\pm}^2 \right]  + 4 r_b \mathrm{Re}(Z_{Tb}^L) Z_{Tb}^{R*} \right\} \,.
\label{ec:GammaT}
\end{align}
Here, $r_x = m_x / m_T$, where $x$ refers to one of the decay products, and the function $\lambda(x,y,z)$ is defined as:
\begin{equation}
\lambda(x,y,z) \equiv (x^4 + y^4 + z^4 - 2 x^2 y^2 
- 2 x^2 z^2 - 2 y^2 z^2) \,,
\end{equation}%
and \begin{eqnarray}
\begin{array}{cc}
Z^L_{Tb}=c_L^d s_L^u e^{- i\phi_u}  +  ( s_L^u{}^2 -  s_R^u{}^2  )\frac{s_L^d}{c_L^u},\\Z^R_{Tb}= \frac{m_b}{ m_T} \left[    c_L^d s_L^u +  (s_R^d{}^2   -  s_L^d{}^2 ) \frac{c_L^u}{s_L^d}     \right]
\end{array}
\end{eqnarray}
\subsection*{Theoretical and Experimental Bounds}

In this section, we outline the constraints used to validate our results.
\begin{itemize}
	\item \textbf{Unitarity} constraints require the $S$-wave component of the various
	(pseudo)scalar-(pseudo)scalar, (pseudo)scalar-gauge boson, and gauge-gauge bosons scatterings to be unitary
	at high energy ~\cite{Kanemura:1993hm}.
	\item \textbf{Perturbativity} constraints impose the following condition on the quartic couplings of the scalar potential: $|\lambda_i|<8\pi$ ($i=1, ...5$)~\cite{Branco:2011iw}.    
	\item \textbf{Vacuum stability} constraints require the potential to be bounded from below and positive in any arbitrary
	direction in the field space, as a consequence, the $\lambda_i$ parameters should satisfy the conditions as~\cite{Barroso:2013awa,Deshpande:1977rw}:
	\begin{align}
	\lambda_1 > 0,\quad\lambda_2>0, \quad\lambda_3>-\sqrt{\lambda_1\lambda_2} ,\nonumber\\ \lambda_3+\lambda_4-|\lambda_5|>-\sqrt{\lambda_1\lambda_2}.\hspace{0.5cm}
	\end{align} 
	\item \textbf{Constraints from EWPOs}, implemented through the oblique parameters\footnote{A comprehensive discussion on EWPO contributions in VLQs can be found in~\cite{Benbrik:2022kpo, Abouabid:2023mbu}}, $S$ and $T$ ~\cite{Grimus:2007if},  require that, for a parameter point of our
	model to be allowed, the corresponding $\chi^2(S^{\mathrm{2HDM\text{-}II}}+S^{\mathrm{VLQ}},~T^{\mathrm{2HDM\text{-}II}}+T^{\mathrm{VLQ}})$ is within 95\% Confidence Level (CL) in matching the global fit results \cite{Molewski:2021ogs}:
	\begin{align}
	S= 0.05& \pm 0.08,\quad T = 0.09 \pm 0.07,\nonumber \\ &\rho_{S,T} = 0.92 \pm 0.11. 
	\end{align}
	Note that unitarity, perturbativity, vacuum stability, as well as $S$ and $T$ constraints, are enforced through the public code  \texttt{2HDMC-1.8.0}\footnote{The code has been adjusted to include new VLQ couplings, along with the integration of analytical expressions for $S_{VLQs}$ and $T_{VLQs}$ outlined in Ref.~\cite{Arhrib:2024tzm}.} \cite{Eriksson:2009ws}.
	\item \textbf{Constraints from the SM-like Higgs-boson properties}  are taken into account by using \texttt{HiggsSignal-3} \cite{Bechtle:2020pkv,Bechtle:2020uwn} via \texttt{HiggsTools-1.2} \cite{Bahl:2022igd}. We require that the relevant quantities (signal strengths, etc.) satisfy $\Delta\chi^2=\chi^2-\chi^2_{\mathrm{min}}$ for these measurements at 95\% CL ($\Delta\chi^2\le6.18$).
	\item\textbf{Constraints from direct searches at colliders}, i.e., LEP, Tevatron, and LHC, are taken at the 95\% CL and are tested using \texttt{HiggsBouns-6}\cite{Bechtle:2008jh,Bechtle:2011sb,Bechtle:2013wla,Bechtle:2015pma} via \texttt{HiggsTools}. Including the most recent searches for neutral and charged scalars.
	
	In addition, the loop contributions of VLQs to $h \to gg$ and $h \to \gamma\gamma$ have been analyzed in our previous study \cite{Arhrib:2024tzm}. These effects are minimal due to the decoupling behavior of VLQs at higher masses and the constraints on the mixing angles ($s^u_{L,R} \sim 0.2$). Consequently, $\mathcal{BR}(h \to gg)$ and $\mathcal{BR}(h \to \gamma\gamma)$ decrease by up to 10\% and 3\%, respectively, primarily due to modifications in the $ht\bar{t}$ coupling, while the direct $hT\bar{T}$ contributions remain negligible. For a detailed discussion of these effects, we refer the reader to Ref.~\cite{Arhrib:2024tzm}.
	
	\item {\bf Constraints from $b\to s\gamma$}:  As established in \cite{Benbrik:2022kpo}, the Type-II 2HDM typically requires the charged Higgs boson mass to exceed 580 GeV to satisfy $b \to s\gamma$ constraint. The introduction of VLQs into the 2HDM can relax this stringent requirement, particularly when larger mixing angles are considered. However, restrictions from EWPOs limit the extent of these mixing angles, permitting charged Higgs masses below 580 GeV. In typical scenarios, the charged Higgs mass remains around 580 GeV for the 2HDM+$T$ singlet and approximately 360 GeV or higher for the 2HDM-II + $TB$ doublet, though the potential for lower masses exists under these relaxed conditions.
	
	\item \textbf{LHC direct search constraints for VLQs}: The current LHC limits primarily target the SM decay modes of VLT, such as $T \to Wb$, $ht$, and $Zt$. In our scenario, however, additional decay channels like $T \to H^\pm b$, $Ht$, and $At$ become relevant, which may influence these limits. To ensure accuracy, we incorporated the latest LHC limits \cite{ATLAS:2022hnn, ATLAS:2022ozf, ATLAS:2022tla, ATLAS:2023bfh, ATLAS:2023ixh, ATLAS:2023pja, ATLAS:2024gyc, ATLAS:2024kgp, ATLAS:2024xdc, ATLAS:2024xne, CMS:2021mku, CMS:2022fck, CMS:2022yxp, CMS:2023agg, CMS:2024bni, CMS:2024qdd, CMS:2024xbc} into our analysis. We implemented a stringent condition where only parameter points meeting the criterion $r={\sigma_{\mathrm{theo}}}/{\sigma_{\mathrm{obs}}^{\mathrm{LHC}}} < 1$ were retained, signifying that points with $r \ge 1$ are excluded at the 95\% CL.	
\end{itemize}

\section{Analysis and Simulation}
\label{sec-A}

\begin{figure}[htbp!]
	\centering
	\includegraphics[height=7.75cm,width=8cm]{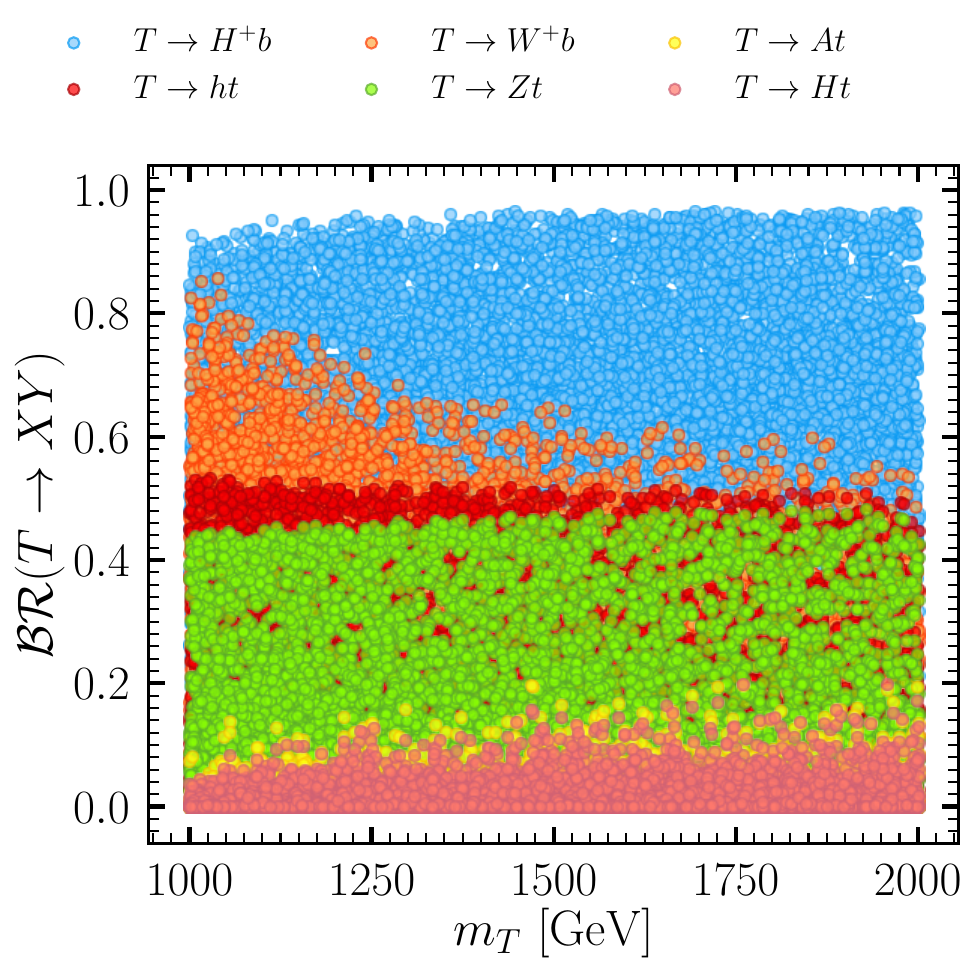}
	\caption{ $\mathcal{BR}(T \to XY)$ as a function of $m_T$, with $XY = H^+ b$ (light blue), $h t$ (red), $Ht$ (salmon), $At$ (yellow), $W^+ b$ (orange), and $Z t$ (green).}
	\label{Fig1}
\end{figure}

This study investigates the 2HDM + $TB$ framework, where the VLT predominantly decays into a charged Higgs boson ($H^+$) and a bottom quark ($T \to H^+ b$). The branching ratio for this decay reaches nearly 100\% for high $m_T$, making it the dominant mode in this scenario. This distinctive feature contrasts with other VLQ representations, where competing decay channels are more prevalent. As shown in Fig.~\ref{Fig1}, the $T \to H^+ b$ mode is the dominant one for large  $m_T$ values ($> 1$ TeV).  Further insights into this behavior can be found in Ref.~\cite{Benbrik:2022kpo}.
\begin{table}[ht!]
	\centering
	\renewcommand{\arraystretch}{0.8}
	\setlength{\tabcolsep}{0.15\columnwidth}
	\begin{tabular}{cc}\Xhline{0.85pt}  \addlinespace[1pt]\Xhline{0.5pt}	
		Parameter  & Range \\
		\Xhline{0.85pt}  		
		$m_h$   & $125.09$ GeV \\
		$m_A$  & [$400$, $800$] GeV \\
		$m_H$  & [$400$, $800$] GeV \\
		$m_{H^\pm}$  & [$400$, $800$] GeV \\
		$t_\beta$ & [$1$, $20$] \\
		$s_{\beta-\alpha}$ & $1$ \\
		$m^2_{12}$ & $m_A^2 s_\beta c_\beta$ \\
		$m_T$   & [$1000$, $2000$] GeV \\	
		$s_L^{u}$  & [$-0.8$, $0.8$] \\
		$s_R^{d}$  & [$-0.8$, $0.8$] \\
	\Xhline{0.5pt}  \addlinespace[1pt]	\Xhline{0.85pt}
	\end{tabular}	
	\caption{Parameter ranges explored for the 2HDM + $TB$ framework.}
	\label{tab:Tab1}
\end{table}

To examine the parameter space comprehensively, we conducted a scan over the ranges listed in Table~\ref{tab:Tab1}. This scan identified four benchmark points (BPs) that satisfy all theoretical and experimental constraints, corresponding to $m_T = 1209.30$, $1452.71$, $1646.60$ and $1877.38$ GeV. The details of these BPs are provided in Table~\ref{tab:Tab2}. These benchmarks serve as representative scenarios where the $T \to H^+ b$ decay dominates, making them optimal for studying the collider signatures of VLT in this framework.

\begin{table}[H]
	\centering
	\renewcommand{\arraystretch}{0.85}
	\begin{adjustbox}{max width=\textwidth}		
		\begin{tabular}{lcccc}
			\Xhline{0.85pt}  \addlinespace[1pt]\Xhline{0.5pt}	
			Parameters &       BP$_1$ &       BP$_2$ &       BP$_3$ & BP$_4$ \\
				\Xhline{0.85pt}
			\multicolumn{5}{c}{(Masses are in GeV)} \\\hline
			$m_h$   & 125.09   & 125.09  &   125.09  & 125.09   \\
			$m_H$ &   640.62 &   640.99 &   655.93 &   651.30 \\
			$m_A$  &  639.27 &   640.51 &   655.27 &   648.64 \\
			$m_{H\pm}$ &   642.49 &   648.62 &   647.40 &   671.88 \\
			$t_\beta$ &    5.51 &     5.04 &     4.22 &     5.22 \\
			$s_{\beta-\alpha}$ &1&1&1&1\\
			$m_T$   & 1209.30 &  1452.71 &  1646.60 &  1877.38 \\
			$m_B$   &  1217.69 &  1468.14 &  1662.59 &  1888.71 \\
			$s^u_L$    & 0.00008 &     0.00659 &    -0.00706 &    -0.00221 \\
			$s^d_L$     & 0.00046 &    -0.00051 &    -0.00045 &    -0.00028 \\
			$s^u_R$     & 0.00056 &     0.05536 &    -0.06717 &    -0.02405 \\
			$s^d_R$   &   0.11718 &    -0.15450 &    -0.15338 &    -0.11194 \\\hline
			\multicolumn{5}{c}{$\mathcal{BR}$ in \%} \\\hline
			${\cal BR}(T\to H^+b)$  &  93.98 & 93.49 & 91.43 & 95.19 \\
			${\cal BR}(H^+\to tb)$  &    95.63 & 96.86 & 98.33 & 96.42 
			\\ 	\Xhline{0.5pt}  \addlinespace[1pt]	\Xhline{0.85pt}
		\end{tabular}
	\end{adjustbox}
	\caption{Benchmark points (BPs) for the 2HDM + $TB$ framework. Masses are in GeV.}\label{tab:Tab2}
\end{table}

\begin{figure*}[htbp!]
	\centering
	\includegraphics[height=4.cm, width=5.5cm]{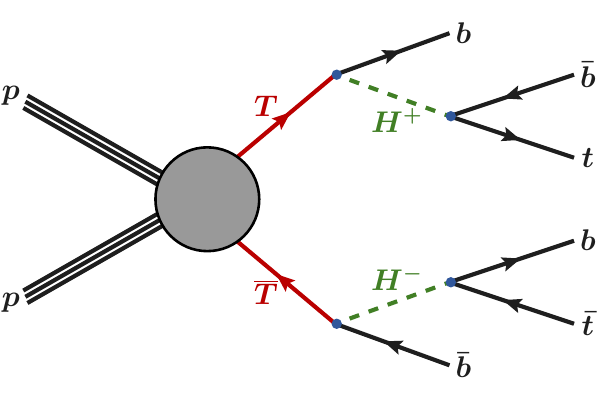}\hspace{1cm}
	\includegraphics[height=4.cm, width=5.5cm]{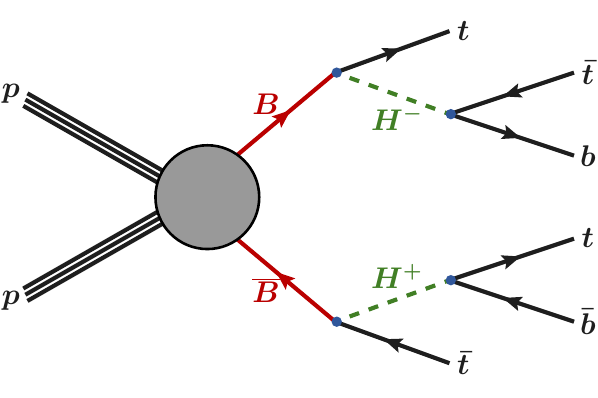}
	\caption{Feynman diagrams for pair production of $T$ (left) and $B$ (right) quarks and their subsequent decays into $4b2t$ and $4t2b$, respectively.}
	\label{diag}
\end{figure*}

The primary signal process is $pp \to T\bar{T} \to 2H^\pm + 2b$, with each $H^\pm$ decaying into $tb$. Subsequent top quark decays lead to a final state of $4b+2t$. The Feynman diagrams for $T$ and $B$ pair production and decay are shown in Fig.~\ref{diag}. While both diagrams yield the same $6b$ final state (after $t \to bW^+$), we focus on $T$ pair production, as $\mathcal{BR}(T \to H^+ b)$ reaches 98\%, compared to the 34\% maximum for $\mathcal{BR}(B \to H^+ t)$~\cite{Arhrib:2024dou}.

We analyzed two tagging scenarios: one requiring at least four $b$-jets (4$b$ analysis) and another requiring five $b$-jets (5$b$ analysis). Event generation was performed using \texttt{MadGraph5\_aMC@NLO} \cite{Alwall:2014hca}, with parton showering and hadronization handled by \texttt{PYTHIA-8.2}~\cite{Sjostrand:2014zea}, and detector effects simulated with \texttt{Delphes-3.4.2}~\cite{deFavereau:2013fsa}. Jet reconstruction utilized the anti-$k_t$ algorithm~\cite{Cacciari:2008gp} with $R = 0.4$ and $p_T (j) > 20$ GeV, while $b$-tagging efficiencies were applied using the \texttt{Delphes CMS} card~\cite{CMS:2012feb}. 

For the parton distribution functions (PDFs), we used the \texttt{NN23LO1} set~\cite{NNPDF:2014otw}. To simulate SM backgrounds with jet multiplicities relevant to our analysis, the MLM matching scheme~\cite{Hoeche:2005vzu} was applied to ensure consistent merging of matrix element and parton shower calculations. Higher-order corrections were incorporated by scaling cross-sections with $K$-factors from the literature (Table~\ref{K-factor}). For the signal, the NNLO $K$-factor was determined to be 1.42 using the \texttt{Top++} package~\cite{Czakon:2011xx}.

\begin{table}[H]
	\centering
	\setlength{\tabcolsep}{5pt}
	\renewcommand{\arraystretch}{1.3}
	\begin{adjustbox}{max width=\textwidth}
		\begin{tabular}{cccccc} 
		\Xhline{0.85pt}  \addlinespace[1pt]\Xhline{0.5pt}	
			Processes & $b\bar{b}jets$ & $b\bar{b}b\bar{b}j $ & $t\bar{t}b\bar{b}$  &  $t\bar{t}t\bar{t}$ & $b\bar{b}b\bar{b}b\bar{b} $\\  
				\Xhline{0.85pt}  
			
			$K$-factor & 1.33\cite{Alwall:2014hca} &1.4\cite{Bevilacqua:2013taa} &1.77\cite{Bevilacqua:2009zn} &1.27\cite{Bevilacqua:2012em} & 2\cite{Papaefstathiou:2019ofh}  \\  
			
		\Xhline{0.5pt}  \addlinespace[1pt]\Xhline{0.85pt}	
		\end{tabular}
	\end{adjustbox}
	\caption{The $K$-factors of the QCD corrections for the background processes.}	\label{K-factor}	
\end{table} 
After event generation, selection cuts were applied to enhance signal significance. Events were required to contain at least four $b$-jets satisfying: 
\begin{itemize} \item $p_T^b > 100$ GeV, $p_T^j > 20$ GeV,
	 \item $|\eta^{b,j}| < 2.5$, 
	  \item $\Delta R(x, y) > 0.4$ for $x, y = j, b$. 
\end{itemize} The SM input parameters used are:
\begin{eqnarray}
&&m_{t}=172.6 ~\mathrm{GeV}, \quad m_{Z}=91.153 ~\mathrm{GeV},\nonumber\\ &&\sin^2(\theta_{W})=0.2226,  \quad\alpha({m_{Z}})= 1/127.934 
\end{eqnarray}

\subsection{4b Analysis}

\begin{figure*}[htbp!]
	\centering
	\includegraphics[height=11cm,width=12cm]{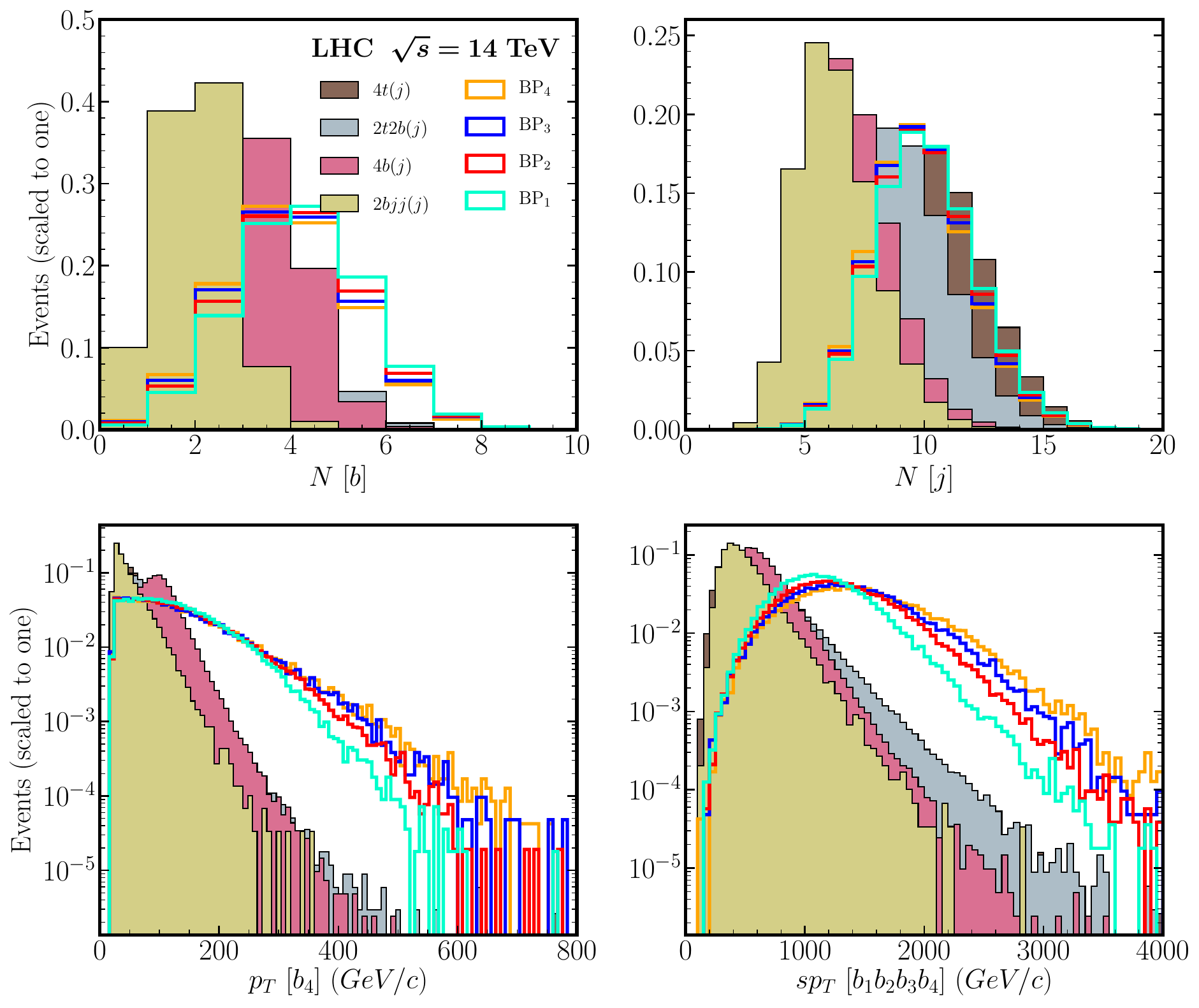}
	\caption{Distributions of $N[b]$, $N[j]$, $p_T(b_4)$, and $sp_T(b_1b_2b_3b_4)$ for the signal benchmarks (BP$_1$, BP$_2$, BP$_3$, BP$_4$) and backgrounds at $\sqrt{s} = 14$ TeV.} 
	\label{Fig4}
\end{figure*}

This section focuses on the primary backgrounds for the signal involving four high-$p_T$ $b$-tagged jets. Key background processes include $2bjj(j)$, $4b(j)$, $2t2b(j)$, and $4t(j)$. Fig~\ref{Fig4} illustrates the distributions of critical variables used to distinguish the signal from the background. The analysis identifies specific regions of phase space where the signal demonstrates significant enhancement over the background, offering valuable insights into the kinematic features of the signal events.

The analyzed observables include the number of $b$-tagged jets ($N[b]$) and jets ($N[j]$), the transverse momentum of the fourth $b$-tagged jet ($p_T[b_4]$), and the total transverse energy of reconstructed $b$-tagged jets ($sp_T$). The selection criteria applied to the signal and background events are summarized in Table~\ref{cutflow1}.

\begin{table}[H]
	\centering
	\setlength{\tabcolsep}{32pt}
	\renewcommand{\arraystretch}{1.0}
	\begin{adjustbox}{max width=\textwidth}
		\begin{tabular}{cc} 
				\Xhline{0.85pt}  \addlinespace[1pt]\Xhline{0.5pt}	
			Cuts & Definition \\  
		\Xhline{0.85pt}  
			Cut 1 & $N(b) \geq 4$ , $N(j) \geq 8$ \\  
			Cut 2 & $sp_T > 1400$ GeV \\  
			Cut 3 & $p_T^{b_4} > 140$ GeV \\
		\Xhline{0.5pt}  \addlinespace[1pt]\Xhline{0.85pt}	
		\end{tabular}
	\end{adjustbox}
	\caption{Selection criteria for signal and background events at $\sqrt{s} = 14$ TeV.}
	\label{cutflow1}
\end{table}

Table~\ref{cutflow2} presents the cut flows for the signal and background events at $\sqrt{s} = 14$ TeV. The selection cuts effectively reduce the background while maintaining high signal efficiency. After applying all cuts, the signal efficiencies are $6.48\%$, $8.83\%$, $9.70\%$, and $9.74\%$ for BP$_1$, BP$_2$, BP$_3$, and BP$_4$, respectively. In contrast, background efficiencies are suppressed to the order of $\mathcal{O}(10^{-4})$.

\begin{table*}[htbp!]
\centering %
\setlength{\tabcolsep}{4.5pt}
\renewcommand{\arraystretch}{1.0}
\begin{adjustbox}{max width=\textwidth}	
	\begin{tabular}{p{1.7cm}<{\centering} p{1cm}<{\centering} p{1cm}<{\centering} p{1cm}<{\centering} p{0.2cm}<{\centering}  p{1.6cm}<{\centering} p{1.6cm}<{\centering} p{1.4cm}<{\centering} p{1.4cm}<{\centering} p{1cm}<{\centering} p{1cm}<{\centering}  p{0cm}<{\centering}p{0cm}<{\centering}}
				\Xhline{0.85pt}  \addlinespace[1pt]\Xhline{0.5pt}	
		\multirow{2}{*}{Cuts}& \multicolumn{4}{c}{Signals }& \multicolumn{4}{c}{\hspace{3cm}Backgrounds}&  \\ \cline{2-5}  \cline{7-10}
		& BP1 & BP2 & BP3& BP4 && $ 2bjets$ & $4bj$ & $2t2bj$ & $4tj$ \\    \cline{1-9} \hline
		Basic& 10.36  &  1.95  &  0.819 & 0.28  & &324075 & 727.74 & 202.56 &  16.60  \\
			\Xhline{0.85pt} 	
		
		Cut1  &  4.41  & 0.760  &  0.31  & 0.093 &&  1005.20  &  32.9  & 29.60  & 3.07  \\
		Cut 2 &   01.00 &  0.272  & 0.13 & 0.045 &&  03.60    & 0.33  & 0.543 & 0.033
		\\
		Cut 3 & 0.67 & 0.168  & 0.08 &   0.026 &&  0.96 & 0.179 & 0.18 & 0.008
		\\
		Efficiency & $6.48\% $ &$ 08.83\% $ & $9.70 \% $&$09.74 \% $  &&  $ 2.93 E^{-6}$  & $ 2.45E^{-4}$    & $ 9.39E^{-4}$ &   $ 4.94E^{-4}$
		\\
			\Xhline{0.5pt}  \addlinespace[1pt]\Xhline{0.85pt}				
	\end{tabular}
\end{adjustbox}
\caption{Cut flow for the signals and backgrounds at $\sqrt{s} = 14$ TeV, with cross sections in fb.}\label{cutflow2}
\end{table*}	
The significance of the signal was evaluated using the median significance approach~\cite{Cowan:2010js}. The discovery significance ($\mathcal{Z}_\mathrm{disc}$) was calculated using the following formulas:
\begin{widetext}
	\begin{eqnarray}
	\mathcal{Z}_\mathrm{disc} &=& \sqrt{2\left[\left(s+b\right)\ln\left(\frac{\left(s+b\right)\left(1+\delta^2b\right)}{b+\delta^2b\left(s+b\right)}\right)-\frac{1}{\delta^2}\ln\left(1+\delta^2\frac{s}{1+\delta^2b}\right)\right]},
	\end{eqnarray}
\end{widetext}
where
\begin{equation*}
x = \sqrt{\left(s+b\right)^2 - 4\delta^2sb^2/\left(1+\delta^2b\right)}.
\end{equation*}
\begin{table*}[htbp!]
	\centering
	\includegraphics[height=5.5cm,width=14cm]{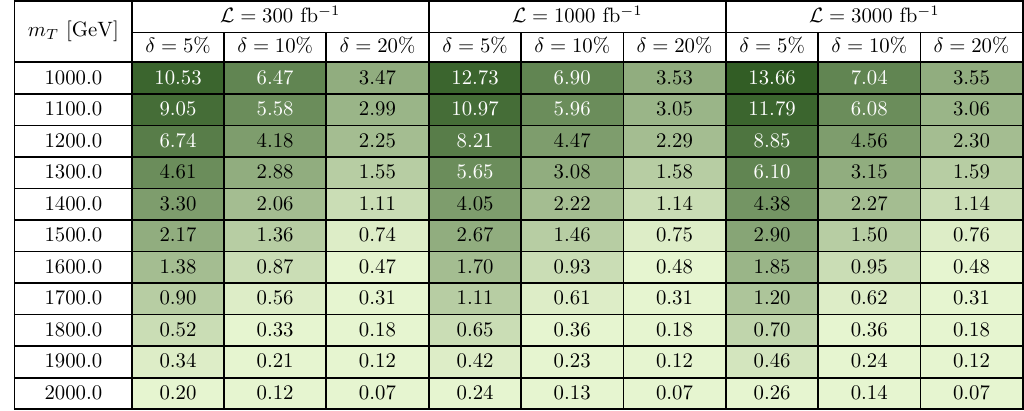}
	\caption{Discovery significance, $\mathcal{Z}_\mathrm{disc}$, presented at varying systematic uncertainties ($\delta$) and integrated luminosities ($\mathcal{L} = 300$ fb$^{-1}$$\mathcal{L} = 1000$ fb$^{-1}$, and $\mathcal{L} = 3000$ fb$^{-1}$). The parameters are fixed as follows: $m_H = 600.26$ GeV, $m_A = 595.24$ GeV, $m_{H^\pm} = 658.07$ GeV, $\tan\beta = 6$, $s^u_R = 0.05$, and $s^d_R = 0.11$. All points depicted are consistent with the discussed constraints.}\label{Tab_dis4b}
\end{table*}

To provide a comprehensive overview of the discovery prospects for VLTs, we extend our analysis to cover the mass range $m_T \in [1000, 2000]$~GeV, building upon the features of the previously discussed benchmark scenarios. For each point, the discovery significance ($\mathcal{Z}_\mathrm{disc}$) is evaluated at $\sqrt{s} = 14$~TeV for integrated luminosities $\mathcal{L} = 300$, 1000, and 3000~fb$^{-1}$, under systematic uncertainties $\delta = 5\%$, 10\%, and 20\%. All points considered are consistent with both theoretical and experimental constraints.

In the baseline case with $m_{H^\pm} \sim 658$~GeV, Table~\ref{Tab_dis4b} summarizes the discovery reach across the full $m_T$ range. At $\mathcal{L} = 300$~fb$^{-1}$, discovery ($\mathcal{Z}_\mathrm{disc} \geq 5$) is achievable up to $m_T = 1200$~GeV for $\delta = 5\%$. For $\delta = 10\%$, the reach reduces to $m_T = 1100$~GeV, while for $\delta = 20\%$, discovery becomes unattainable, and only evidence-level sensitivity is retained below $m_T \sim 1000$~GeV. At 1000~fb$^{-1}$, the reach improves significantly, extending to $m_T = 1300$~GeV for $\delta = 5\%$, and to $m_T = 1100$~GeV for $\delta = 10\%$, while remaining below threshold for $\delta = 20\%$. The discovery reach at 3000~fb$^{-1}$ shows only moderate improvement, saturating near $m_T = 1300$~GeV for $\delta = 5\%$.

\begin{figure*}[htbp!]
	\centering
	\includegraphics[height=6.25cm,width=16cm]{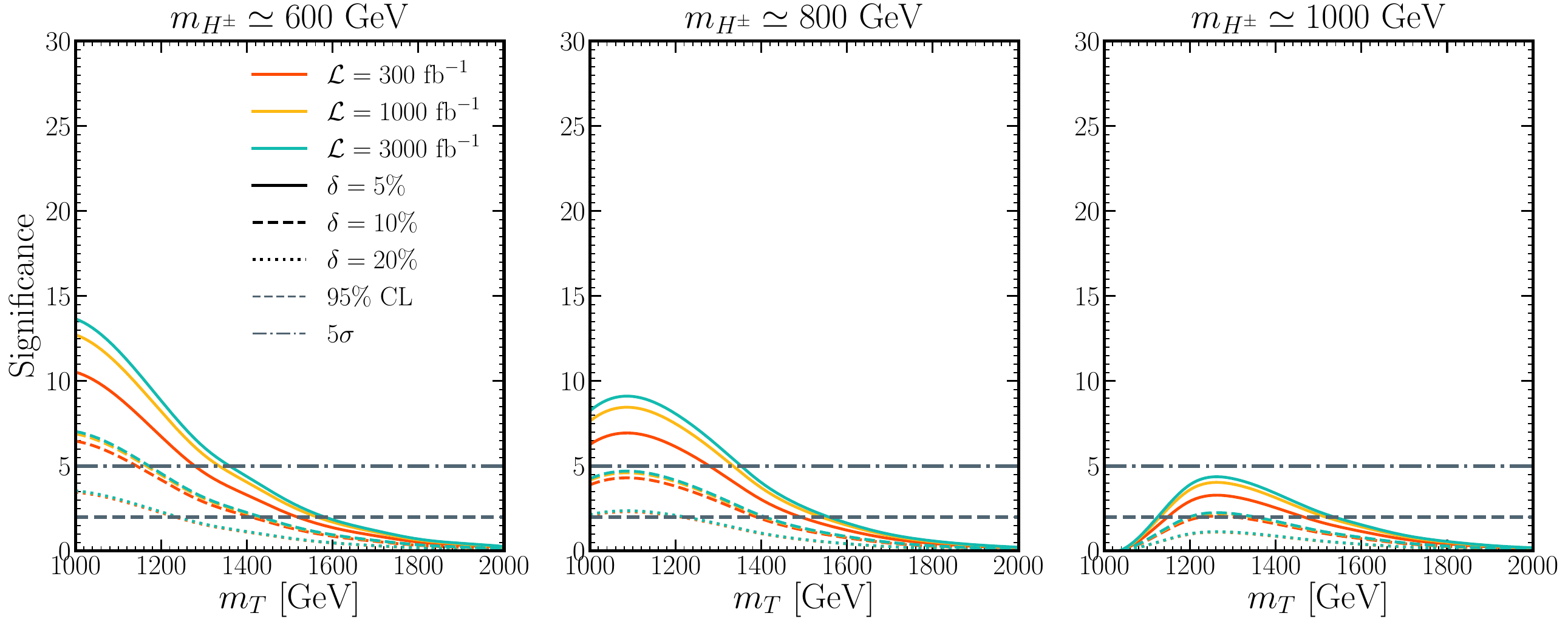}
	\caption{Significance as a function of $m_T$ for three benchmark charged Higgs masses: $m_{H^\pm} = 600$~GeV (left), 800~GeV (middle), and 1000~GeV (right), shown for different systematic uncertainties ($\delta$) and integrated luminosities $\mathcal{L} = 300$, 1000, and 3000~fb$^{-1}$. The parameters are fixed to $m_H \simeq m_A \simeq m_{H^\pm}$, $\tan\beta = 5$, $s_R^u = 0.05$, and $s_R^d = 0.11$. All points depicted are consistent with the discussed constraints.}\label{Hp4b}
\end{figure*}

Fig.~\ref{Hp4b} complements Table~\ref{Tab_dis4b} by illustrating the discovery significance as a function of $m_T$ for three representative charged Higgs masses: $m_{H^\pm} = 600$, 800, and 1000~GeV. The horizontal grey lines indicate the $5\sigma$ discovery and 95\% CL exclusion thresholds. For $m_{H^\pm} = 600$~GeV, discovery-level significance is attainable up to $m_T \sim 1300$~GeV at 300~fb$^{-1}$ for $\delta = 5\%$, and up to 1200~GeV for $\delta = 10\%$. For $\delta = 20\%$, discovery is not possible. At 1000 and 3000~fb$^{-1}$, the reach improves to $m_T \sim 1400$~GeV for $\delta = 5\%$, though higher systematic uncertainties continue to limit the sensitivity.

As $m_{H^\pm}$ increases, the reduced $\text{BR}(T \to H^+ b)$ leads to a significant drop in sensitivity. For $m_{H^\pm} = 800$~GeV, discovery is only possible for $\delta = 5\%$, with a more limited $m_T$ range compared to the $m_{H^\pm} = 600$~GeV case. For $\delta \geq 10\%$, discovery remains unattainable across the full range. In the scenario with $m_{H^\pm} = 1000$~GeV, even with $\mathcal{L} = 3000$~fb$^{-1}$ and $\delta = 5\%$, the significance stays below the discovery threshold. These findings reinforce the conclusion that charged Higgs masses below 1~TeV provide the most promising conditions for discovering  a vector-like top partner in this framework at the LHC.

\subsection{5b Analysis}

The 5$b$ analysis focuses on irreducible backgrounds involving events with five high transverse momentum ($p_T$) $b$-tagged jets. The dominant backgrounds include $2bjets$, $4bj$, $2t2bj$, and $6b$, where misidentifications of jets as $b$-jets contribute significantly. Backgrounds such as $4t2b$, $4b2t$, and $2tjets$ are excluded due to their negligible cross-sections.
\begin{figure}[H]
	\centering
	\includegraphics[height=6.25cm,width=7.75cm]{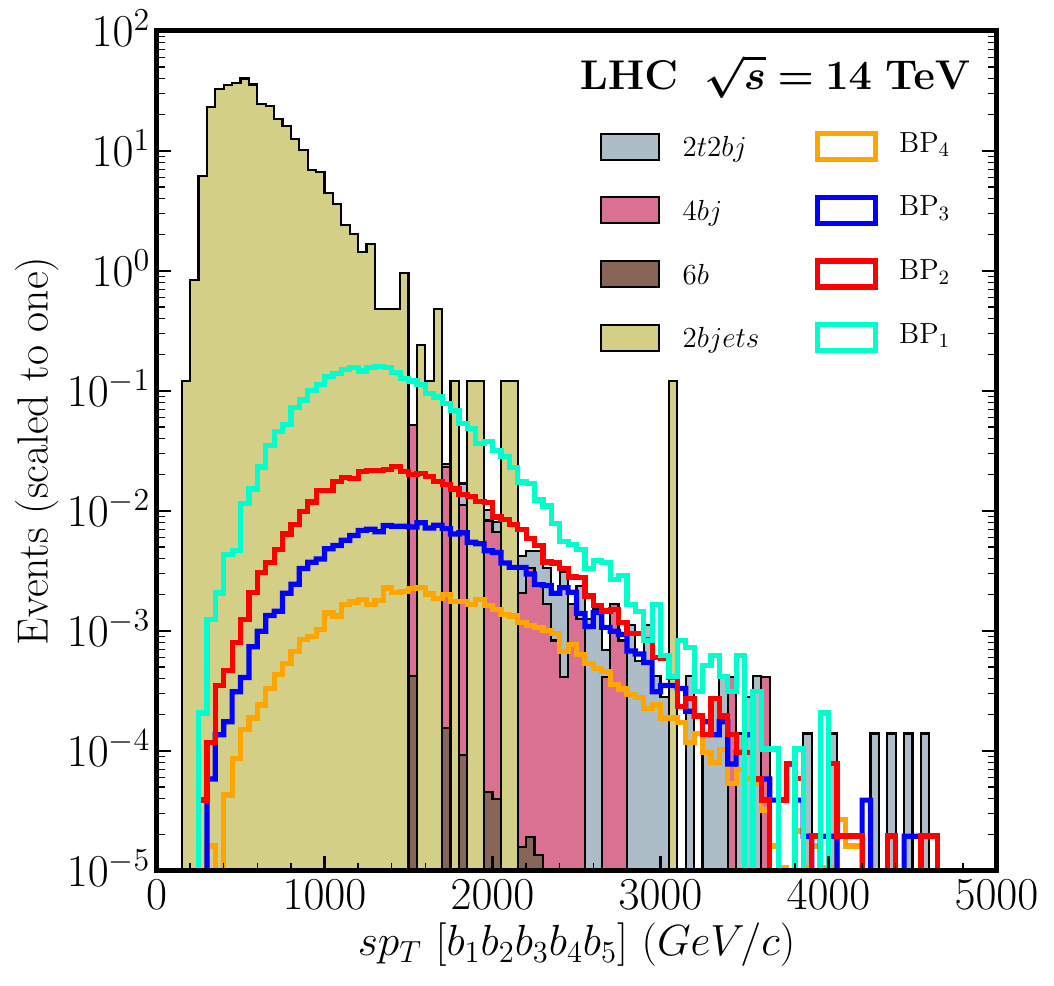}
	\caption{$sp_T$ distribution for the signal BPs and backgrounds in the $5b$ analysis.}\label{Fig7}
\end{figure}
Fig~\ref{Fig7} presents the $sp_T$ distributions for both signal and background events. To maximize signal significance, a series of optimized cuts is applied, as summarized in Table~\ref{cutflow3}.

The cut flow applied to the signal and background events at $\sqrt{s} = 14$ TeV is detailed in Table~\ref{cutflow4}. The results highlight the efficacy of these cuts in suppressing background while retaining a significant fraction of the signal. After applying the cuts, the total efficiencies for the signal reach $7.09\%$, $9.17\%$, $9.75\%$, and $9.73\%$ for BP$_1$, BP$_2$, BP$_3$, and BP$_4$, respectively. Conversely, background efficiencies are significantly reduced, with cumulative values around 1\%.
\begin{table}[H]
	\centering
	\setlength{\tabcolsep}{32pt}
	\renewcommand{\arraystretch}{1.0}
	\begin{adjustbox}{max width=\textwidth}
		\begin{tabular}{cc} 
				\Xhline{0.85pt}  \addlinespace[1pt]\Xhline{0.5pt}	
			Cuts & Definition \\  
				\Xhline{0.85pt}  
			Cut 1 & $N(b) \geq 5$, $N(j) \geq 8$ \\  
			Cut 2 & $sp_T > 1500 \, \text{GeV}$ \\
			Cut 3 & $p_{T}^{b_{4}} > 120 \, \text{GeV}$ \\  
			\Xhline{0.5pt}  \addlinespace[1pt]\Xhline{0.85pt}	
		\end{tabular}
	\end{adjustbox}
	\caption{Selection criteria for the 5$b$ analysis at $\sqrt{s} = 14$ TeV.}	\label{cutflow3}	
\end{table}

\begin{table*}[htbp!]
	\centering %
	\setlength{\tabcolsep}{4.5pt}
	\renewcommand{\arraystretch}{1.0}
	\begin{adjustbox}{max width=\textwidth}	
		\begin{tabular}{p{1.7cm}<{\centering} p{1cm}<{\centering} p{1cm}<{\centering} p{1cm}<{\centering} p{0.2cm}<{\centering}  p{1.6cm}<{\centering} p{1.6cm}<{\centering} p{1.4cm}<{\centering} p{1.4cm}<{\centering} p{1cm}<{\centering} p{1cm}<{\centering}  p{0cm}<{\centering}p{0cm}<{\centering}}
		\Xhline{0.85pt}  \addlinespace[1pt]\Xhline{0.5pt}	
			\multirow{2}{*}{Cuts}& \multicolumn{4}{c}{Signals }& \multicolumn{4}{c}{\hspace{3cm}Backgrounds}&  \\ \cline{2-5}  \cline{7-10}
			& BP1 & BP2 & BP3& BP4 && $ 2bjets$ & $4bj$ & $2t2bj$ & $6b$ \\    \cline{1-9} \hline
			Basic&10.36  &  1.95  &  0.819 & 0.28  & &324075 & 727.74 & 202.56 &  0.146 \\
				\Xhline{0.85pt}  
			
			Cut1  &  2.76  & 0.467  &  0.179  & 0.0545 &&  241.7  &  16.4  & 10.21 & 0.0406  \\
			Cut 2 &   0.841 & 0.21  & 0.096 & 0.032 &&  0.842    & 0.18  & 0.28 & 0.00151
			\\
			Cut 3 & 0.734 & 0.179 & 0.0798 &   0.026 &&  0.722 & 0.145 & 0.178 & 0.0013
			\\
			Efficiency & $7.09\% $ &$ 9.17\% $ & $9.75 \% $&$09.73 \% $  &&  $ 2.23 E^{-6}$  & $ 1.99E^{-4}$    & $ 8.81E^{-4}$ &   $ 9.40 E^{-3}$
			\\
					\Xhline{0.5pt}  \addlinespace[1pt]\Xhline{0.85pt}					
		\end{tabular}
	\end{adjustbox}
	\caption{Cut flow of cross sections (in fb) for the signal and SM backgrounds at $\sqrt{s} = 14$ TeV.}
\label{cutflow4}
\end{table*}	

The corresponding discovery prospects for VLTs in the 5$b$ final state are summarized in Table~\ref{Tab_dis5b}, assuming $m_{H^\pm} \sim 658$~GeV. The analysis is performed at $\sqrt{s} = 14$~TeV for integrated luminosities $\mathcal{L} = 300$, 1000, and 3000~fb$^{-1}$ and systematic uncertainties $\delta = 5\%$, 10\%, and 20\%, in direct parallel to the 4$b$ case. The additional $b$-tag improves background rejection, leading to a modest but consistent gain in sensitivity across the full $m_T$ range.

At $\mathcal{L} = 300$~fb$^{-1}$ and $\delta = 5\%$, discovery-level significance ($\mathcal{Z}_\mathrm{disc} \geq 5$) is achievable for $m_T \leq 1300$~GeV an improvement of approximately 100~GeV over the 4$b$ final state. For $\delta = 10\%$, the reach is reduced to 1200~GeV, while discovery is no longer possible for $\delta = 20\%$, although evidence-level significance remains up to $m_T \sim 1100$~GeV. At higher luminosities, the discovery potential saturates near $m_T \sim 1400$~GeV for $\delta = 5\%$, emphasizing the dominant role of systematic uncertainties in the HL-LHC regime.

\begin{figure*}[htbp!]
	\centering
	\includegraphics[height=6.25cm,width=16cm]{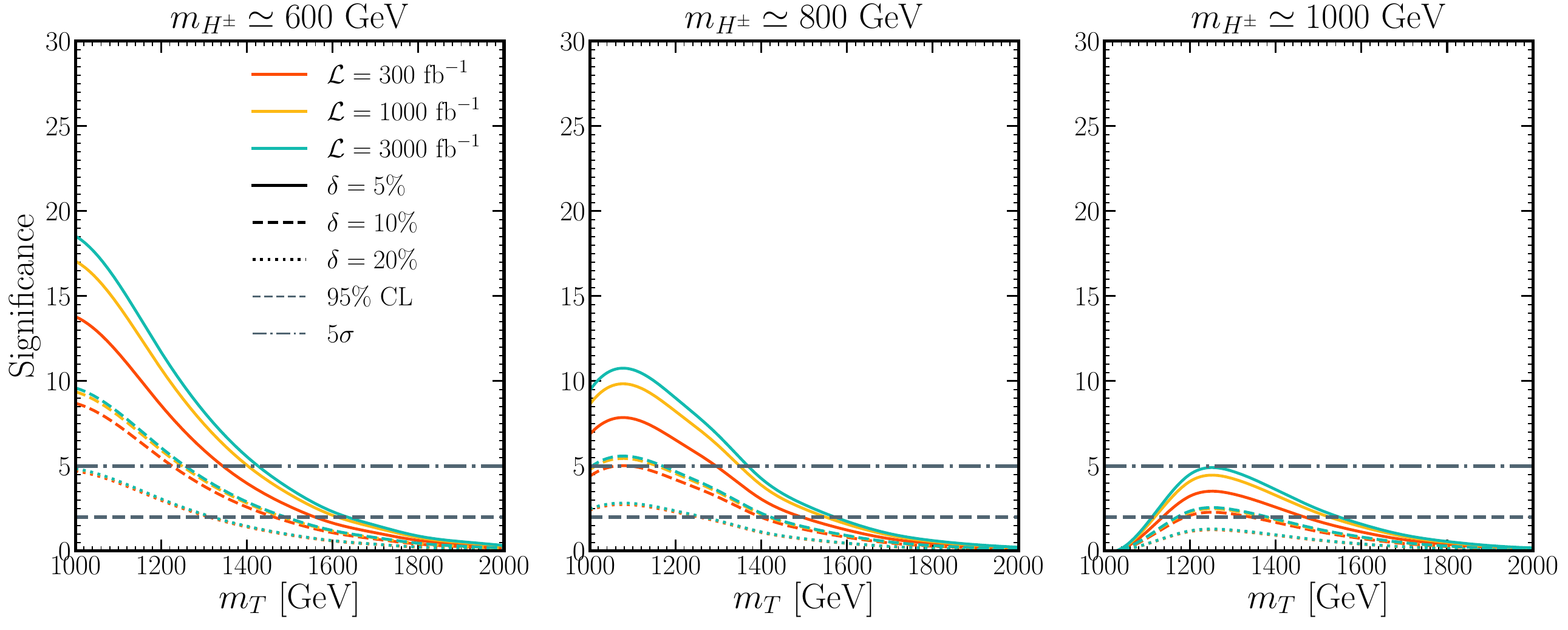}
\caption{Same as Fig.~\ref{Hp4b}, but for the 5$b$ analysis.}\label{Hp5b}
\end{figure*}

Fig.~\ref{Hp5b} complements Table~\ref{Tab_dis5b} by illustrating the significance as a function of $m_T$ for three benchmark charged Higgs masses: $m_{H^\pm} = 600$, 800, and 1000~GeV. Horizontal grey lines denote the $5\sigma$ discovery and 95\% CL exclusion thresholds. The overall behavior closely follows that observed in the 4$b$ scenario, with a slight enhancement in sensitivity due to the additional $b$-tag. 

For $m_{H^\pm} = 600$~GeV, discovery-level significance is achieved up to $m_T \sim 1300$~GeV at $\mathcal{L} = 300$~fb$^{-1}$ and extends to $m_T \sim 1400$~GeV at higher luminosities, provided $\delta = 5\%$. For $m_{H^\pm} = 800$~GeV, the discovery region persists but contracts in $m_T$, reflecting the impact of reduced $\text{BR}(T \to H^+ b)$. In the case of $m_{H^\pm} = 1000$~GeV, discovery is not attainable across the full range of luminosities and $\delta$ values, even under optimal conditions.

\begin{table*}[htbp!]
	\centering
	\includegraphics[height=5.5cm,width=14cm]{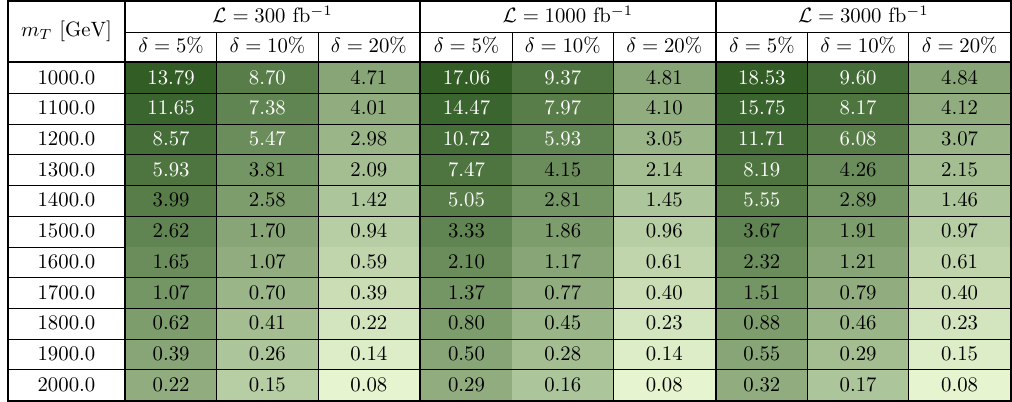}
	\caption{Discovery significances for the 5$b$ analysis, analogous to Table~\ref{Tab_dis4b}.}\label{Tab_dis5b}
\end{table*}

\section{DISCUSSION AND SUMMARY}\label{sec:conclusion}

We explored the discovery prospects of vector-like top partner (VLT) within the 2HDM-II extended by VLQ doublet $TB$ , focusing on the pair production process $pp \to T\bar{T} \to bH^+\, \bar{b}H^- \to b(tb)\, \bar{b}(\bar{t}b)$ at the LHC. The resulting final states are characterized by high $b$-jet multiplicities, analyzed under two complementary strategies: the 4$b$ and 5$b$ final state analyses, requiring at least four and five $b$-tagged jets, respectively.

Our results show that the 5$b$ analysis provides enhanced sensitivity compared to the 4$b$ case, primarily due to improved background suppression enabled by the additional $b$-tag. For a representative charged Higgs mass $m_{H^\pm} \sim 660$~GeV, and under optimistic conditions of $\delta = 5\%$, the 4$b$ (5$b$) analysis enables $5\sigma$ discovery up to $m_T = 1200$ (1300)~GeV at $\mathcal{L} = 300$~fb$^{-1}$. At $\mathcal{L} = 1000$~fb$^{-1}$, the 5$b$ channel extends the reach to $m_T \sim 1400$~GeV, while the gain at 3000~fb$^{-1}$ remains marginal, underscoring the dominant role of systematic uncertainties in the high-luminosity regime.

The impact of increasing systematic uncertainties is significant. For $\delta = 10\%$, the discovery reach drops to $m_T \leq 1100$~GeV (4$b$) and $m_T \leq 1200$~GeV (5$b$) at 300~fb$^{-1}$. At $\delta = 20\%$, neither channel achieves discovery beyond $m_T \sim 1000$~GeV, even at the highest luminosity considered.

We also evaluated the impact of the charged Higgs mass on the discovery reach. For $m_{H^\pm} = 600$~GeV, discovery-level significance is attained up to $m_T \sim 1300$~GeV at $\mathcal{L} = 300$~fb$^{-1}$, and extends to $m_T \sim 1400$~GeV at higher luminosities, assuming $\delta = 5\%$. As $m_{H^\pm}$ increases to 800~GeV, the discovery reach remains viable but narrows, with a reduced $m_T$ window due to the suppression of $\text{BR}(T \to H^+ b)$. In contrast, for $m_{H^\pm} = 1000$~GeV, the branching ratio becomes too suppressed to yield discovery across the entire explored parameter space. Even under the most favorable conditions $\delta = 5\%$ and $\mathcal{L} = 3000$~fb$^{-1}$ the significance remains below the $5\sigma$ threshold, and only evidence-level sensitivity is achieved.

In summary, the 2HDM-II + VLQ framework offers a well-motivated avenue for probing new physics signatures at the LHC. Final states with high $b$-jet multiplicities from $T\bar{T}$ production and subsequent decays into charged Higgs bosons provide clean and distinctive discovery channels. The 5$b$ final state delivers consistent improvements over the 4$b$ scenario, particularly at higher VLT masses. However, the analysis highlights the necessity of controlling systematic uncertainties and achieving high integrated luminosities to fully exploit the discovery potential of such models in future LHC runs.
\section*{ACKNOWLEDGMENTS}
R. B. is supported in part by the PIFI Grant No. 110200EZ52. M. Boukidi acknowledges the support of Narodowe Centrum Nauki under OPUS Grant No. 2023/49/B/ST2/03862.

\bibliography{main1}
\bibliographystyle{JHEP}
\end{document}